# CDF Central Preshower and Crack Detector Upgrade

[1]A. Artikov, [1]J. Boudagov, [1]D. Chokheli, [2]G. Drake, [3,4]M. Gallinaro, [5]M. Giunta,
[2]J. Grudzinski, [6]J. Huston, [7]M. Iori, [8]D. Kim, [9]M. Kim, [9]N. Kimura, [2]S. Kuhlmann,
[3]S. Lami, [6]R. Miller, [9]K. Nakamura, [2]L. Nodulman, [7]A. Penzo, [9]K. Sato, [8]J. Suh,
[5]N. Turini, [9]F. Ukegawa, [9]Y. Yamada

[1]*Joint Institue for Nuclear Research, RU-141980, Dubna, Russia*
[2]*Argonne National Laboratory, Argonne, Illinois 60439*
[3]*The Rockefeller University, New York, New York 10021*
[4]*Present address: LIP, 1000-149 Lisbon, Portugal*
[5]*Istituto Nazionale di Fisica Nucleare Pisa, Universities of Pisa*
*Siena and Scuola Normale Superiore, I-56127 Pisa, Italy*
[6]*Michigan State University, East Lansing, Michigan 48824*
[7]*Istituto Nazionale di Fisica Nucleare, Sezione di Roma 1, University of Rome "La Sapienza," I-00185 Roma, Italy*
[8]*Center for High Energy Physics: Kyungpook National University, Taegu 702-701;*
*Seoul National University, Seoul 151-742; and SungKyunKwan University, Suwon 440-746; Korea*
[9]*University of Tsukuba, Tsukuba, Ibaraki 305, Japan*

**Abstract**

The CDF Central Preshower and Crack Detector Upgrade consist of scintillator tiles with embedded wavelength-shifting fibers, clear-fiber optical cables, and multi-anode photomultiplier readout. A description of the detector design, test results from R&D studies, and construction phase are reported. The upgrade was installed late in 2004, and a large amount of proton-antiproton collider data has been collected since then. Detector studies using those data are also discussed.

# 1. Overview

The Collider Detector at Fermilab (CDF) experiment[1], from 1992 to 2004, used gas proportional chambers in the central calorimeter to improve the identification of electrons and photons. As the Collider instantaneous luminosity increases, the number of overlapping proton-antiproton collisions increases, together with the detector occupancy from background particles. The relatively slow gas detectors were expected to suffer an additional 30-40% channel occupancy increase from these overlapping collisions as the instantaneous luminosity is expected to grow in the next few years. In order to improve electron and photon identification in this environment, the CDF *Central Preshower* (CPR) and *Central Crack* (CCR) detectors have been replaced by faster scintillator counters read out by optical fibers (similar to the CDF end-plug calorimeter upgrade[2]) and multi-anode photomultiplier tubes[3], the CPR2 project[4]. The new preshower detector also has a better segmentation which will help reducing the detector occupancy and improve jet energy resolution.

CDF has two calorimeter barrels along the beam line, 48 CPR and 48 CCR modules were installed as part of the upgrade project. Each calorimeter barrel is split at the top and bottom into two arches of 12 detectors each, which can be independently moved for installation. Figure 1.1 shows a picture of one calorimeter arch that has been moved away from the beam line with the old CPR detectors on the inner face. The CCR detectors are positioned behind a 8.5 radiation length tungsten bar which converts high energy electrons and photons before they escape detection in the gaps between the calorimeter modules.

Figure 1.2 provides another view of the inside of the calorimeter arch during the installation of the upgraded CPR/CCR detectors. This picture shows the front face of three calorimeter wedges. The top wedge has an upgraded CPR module installed, while the middle wedge shows the bare EM calorimeter face, and the bottom wedge shows the old preshower detector before removal. The CCR detectors are installed behind the tungsten bars (visible in the picture) which cover the gaps between the wedges.

The preshower detector is designed to efficiently observe minimum-ionizing particles (MIPs), therefore the specification was such that the light yield from every tile of at least five photoelectrons/MIP after the full optical path was required. To meet this requirement, due to typical variations in scintillator yield, optical path coupling, optical fiber attenuation, as well as detector aging effects, a more stringent specification was a minimum tile yield of 10 photoelectrons/MIP to allow for a 50% fluctuation in the worst tiles.

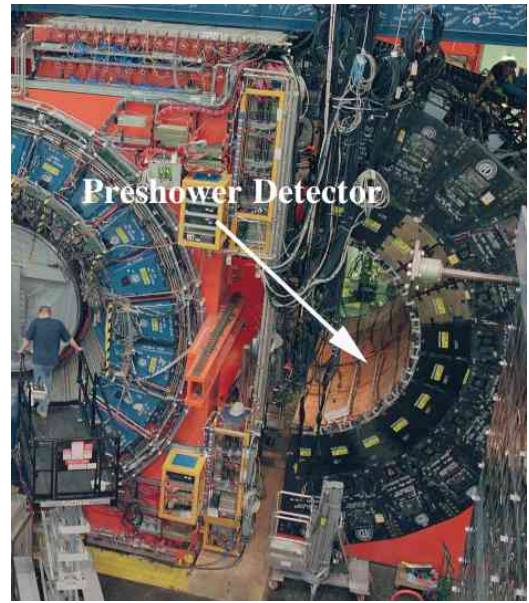

Figure 1.1: Picture of a CDF calorimeter arch moved away from the beam line. The inner surface of the arch is where the CPR and CCR detectors are installed.

Due to the lack of space near the detectors to place photomultipliers (PMTs), it was necessary to bring the light out to the back of the calorimeter wedges with optical fiber cables. Some tiles are more than 6m from the PMT, which led to several design decisions. Since the attenuation loss in WLS fibers alone would have been too large to use them in the complete optical path, the WLS fibers were spliced to clear fibers immediately after exiting the scintillator. In order to simplify the installation procedure the optical path was split at the module's exit, and plastic optical connectors were placed at the end of the detector units, whereas separate clear fiber cables completed the optical paths to the PMTs. The combined loss of light due to attenuation and optical connections was measured to be 60%, hence a good light yield from the scintillator tile itself was needed to meet the specifications. This led to the use of a 2 cm thick scintillator (the thickest possible scintillator in the allowed space), and an extensive R&D program to optimize the

WLS fiber pattern inside the tile for the best light yield.

The CCR detector is not designed to observe MIPs, since it sits behind the tungsten bar and detects electromagnetic showers with hundreds of secondary particles. The available space limits the scintillator thickness to 5 mm, which is what was used in the upgrade. The optical paths of the CPR and CCR detectors are very similar.

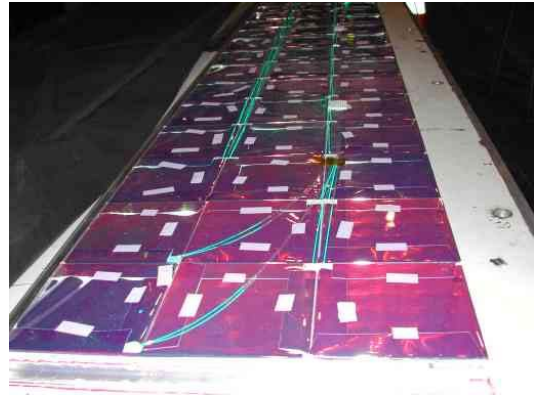

Figure 1.3: Picture of a CPR module before the top cover was installed, showing the individually wrapped tiles and fibers exiting them.

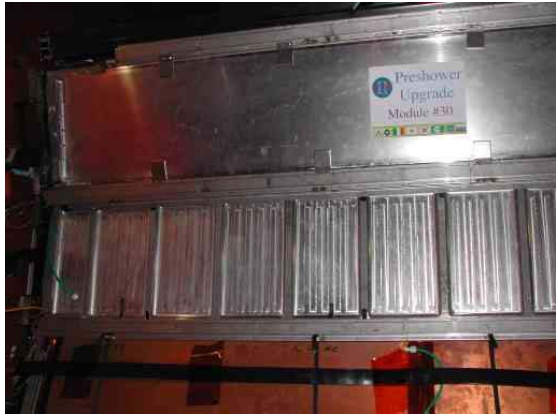

Figure 1.2: Picture of the front face of the inner calorimeter arch, showing three wedges. The middle wedge is bare, showing the front of the electromagnetic calorimeter. The bottom wedge has the old gas preshower detector which was replaced; the top wedge has the new preshower detector.

The detector segmentation was mostly determined by the desire to reuse the existing electronics infrastructure for the new detectors. Of the existing 64 channels per calorimeter wedge, 10 channels were assigned to the CCR detector and 54 to the CPR detector. Based on experience in electron and photon identification, it was decided that six preshower detector tiles covering a single calorimeter tower, three segments in azimuth and two segments in rapidity, was the optimal layout. These choices, combined with the desire to cover as much of the fiducial region as possible, determined the tile sizes of the preshower and crack detectors. The preshower tile dimension (Figure 1.3) is 125x125x20 mm$^3$, while the crack tile dimension is 225x50x5 mm$^3$. For the CPR, a one mm diameter WLS fiber lays in a groove obtained by machining the tile surface, which is then spliced at the tile's exit to a clear fiber, transporting the light to the optical connector placed on the edge of the module (Figure 1.3). In the CCR detector, a straight fiber is inserted in a groove in the center of the tile.

## 2. Preshower Scintillator and Fiber R&D

A cosmic ray test of individual tiles inside a light-sealed box was performed in order to compare different scintillators (Bicron 408[5] and Dubna[1]), multi-clad WLS and clear fibers (Kuraray[6] and Pol.Hi.Tech[7]), and reflectors (Aluminum foil, Tyvek paper, 3M VM2002 reflector), and to optimize the design of the groove in the tile.

Dubna and Bicron tiles were compared and yielded similar light yield response within 5%. The less expensive Dubna scintillator was therefore chosen in the final design. The most probable value of a fit to a Landau distribution indicated the number of pe/MIP collected. Both response uniformity (better than 3%) and light yield of tiles were found equal within uncertainties for two kinds of groove path, circular or sigma ($\sigma$), 5 mm away from the tile edges and with a maximum curvature radius of 5 cm. A spiral $\sigma$-groove was chosen, as discussed in more detail later in this section.

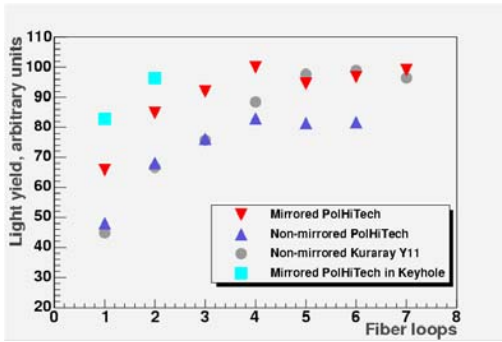

Figure 2.1: Number of photoelectrons/MIP as a function of the number of fiber loops inside the tile for three different fiber types.

The optical coupling between the fiber and the groove was studied in a tile with four loops of WLS fiber inserted in a square cross-section groove. Similar results were obtained with either Bicron BC-600 glue or BC-630 grease, with an increase of about 40% in light yield with respect to air coupling. Figure 2.1 shows the result of the study on the optimal number of fiber loops into a $\sigma$-groove with a square cross section and tiles wrapped with Tyvek paper. The light yield of Pol.Hi.Tech fibers reaches a plateau at N=4 loops, when the increased light yield seems to be compensated by the attenuation of the longer fiber. For a Kuraray Y11(250) fiber, whose attenuation length is about 50 cm longer than Pol.Hi.Tech or about one extra loop, the plateau is reached at N=5 loops. Concerns about glue damaging fibers prompted a study of the response of different glued tiles over a period of 9 months. No time decay was observed, as shown in Figure 2.2.

---

[1] Provided by the JINR Dubna group.

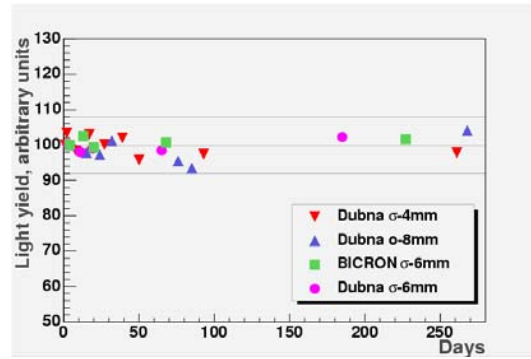

Figure 2.2: Time stability response over the nine months of the tiles tested in our studies.

The glue in the groove served two purposes, increasing the light yield and holding the fiber in place, inside the groove. However, the gluing of every tile would have required significant assembly time, hence a study ensued of an alternative method of embedding a 1-loop WLS fiber into a tile with a keyhole-shaped cross section with air coupling, as shown in Figure 2.3, which was previously used for the CDF Plug Shower Max detector[8].

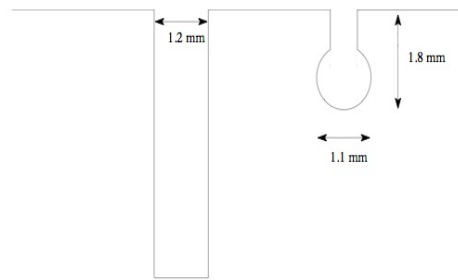

Figure 2.3: The two groove shapes tested for the CPR detector upgrade, described in the text.

Two different groove cross sectional shapes were compared: the original baseline design based on a square cross section, 6 mm deep, loaded with 4 or 5 glued loops of WLS fiber, and the alternative keyhole groove (1.8 mm deep) where no glue is needed to hold the fiber. Other advantages in using the keyhole groove shape are the uniformity in tile preparation and, overall, a better light yield per fiber unit length, due to the keyhole cross sectional design which functions as an extra cladding to better trap the light lost by the fiber.

Figure 2.4 displays test results of the different groove types, performed using Tyvek wrapping and Pol.Hi.Tech mirrored WLS fibers. A 2-loop spiral keyhole groove and no glue yielded a similar number of pe/MIP as the baseline design with four loops and a fiber glued in a square groove. The 2-loop spiral gave 15% more light than a single loop, and was chosen for the final design (Figure 2.5). During the production assembly, the six tiles farthest from the module edge were glued in order to compensate for the effect of a longer fiber attenuation and provide a more uniform response across the module.

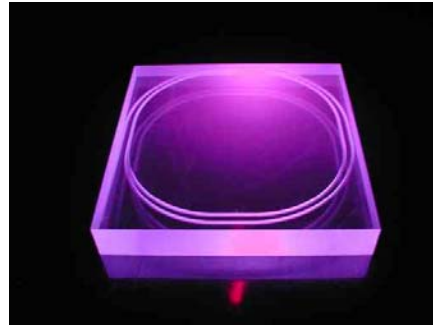

Figure 2.5: Photograph of a production tile under a UV lamp.

The final R&D study was the attenuation of the clear optical fibers. This was performed with a single tile in a cosmic-ray test stand, which was sampled by systematically cutting a fiber shorter to measure the attenuation length. The result for the PolHiTech fibers was an attenuation length of 7.3m, sufficient for the detector needs (Figure 2.6).

| # | Fiber | Mirrored | Reflector | Pe/MIP |
|---|---|---|---|---|
| 1 | Pol.Hi.Tech | No | Tyvek | 24.1 |
| 2 | Pol.Hi.Tech | Yes | Tyvek | 29.4 |
| 3 | Kuraray Y11(250) | No | 3M | 28.1 |
| 4 | Kuraray Y11(250) | Yes | Tyvek | 30.5 |
| 5 | Kuraray Y11(350) | No | Tyvek | 28.7 |
| 6 | Kuraray Y11(350) | No | 3M | 32.9 |
| 7 | Kuraray Y11(350) | Yes | 3M | 38.2 |
| 8 | Kuraray Y11(350)+Glue | Yes | 3M | 43.9 |

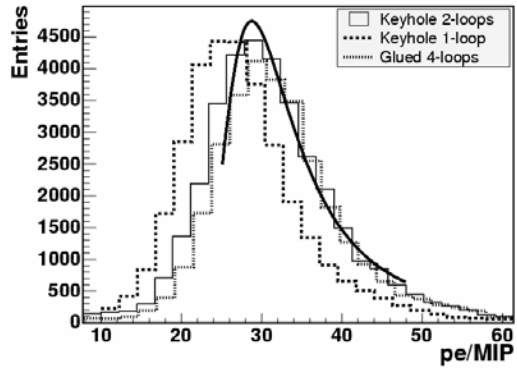

Figure 2.4: The light yield from a 2-loop spiral keyhole groove with air coupling (solid histogram and fit) is compared to the light yield from glued fibers with 4 loops in a square groove (dash-dotted histogram), and to a 1-loop spiral keyhole with air coupling (dashed histogram).

Table 2.1: Number of photoelectrons/MIP for different WLS fibers into a 2-loop spiral keyhole groove.

Other factors affecting light yield were studied. These included the reflective wrap material, the aluminized mirror on the fiber, fiber manufacturer, fiber dopants, and optical coupling (air or grease). The results are summarized in table 2.1. As mentioned above, 2/3 of the tiles, those closest to the PMT, used configuration #7 in the table, and 1/3 of the tiles, those farthest from the PMT, used configuration #8 in order to provide a more uniform response.

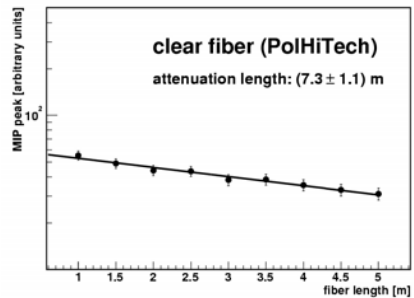

Figure 2.6: Measurement of the attenuation length of fibers used for the optical cables.

## 3. Mechanical Design

The CPR and CCR module lengths and widths were set to maximize the fiducial coverage while safely fitting within the allowed space. The scintillator thicknesses were also maximized within the allowed space to provide the best possible light yield. The CPR module is 2425 x 383 x 26.8 mm, while the CCR module is 2279 x 54.3 x 7.1 mm.

Figures 3.1 and 3.2 show the details of the assembled CPR module. The modules are composed of the scintillator tiles (instrumented with WLS fiber) enclosed within an aluminum case. The case addresses several of the constraints by providing structural rigidity, protection for the fibers and tiles, and light tightness. The aluminum case is constructed of a top and bottom sheet (2.5mm thick) coupled together through aluminum U-channels on the side. The top and bottom skins are bonded using a fast-setting two-part epoxy (3M brand DP-810). Figure 3.3 shows the details of the CCR module, which is much smaller and lighter than the CPR but constructed with similar techniques.

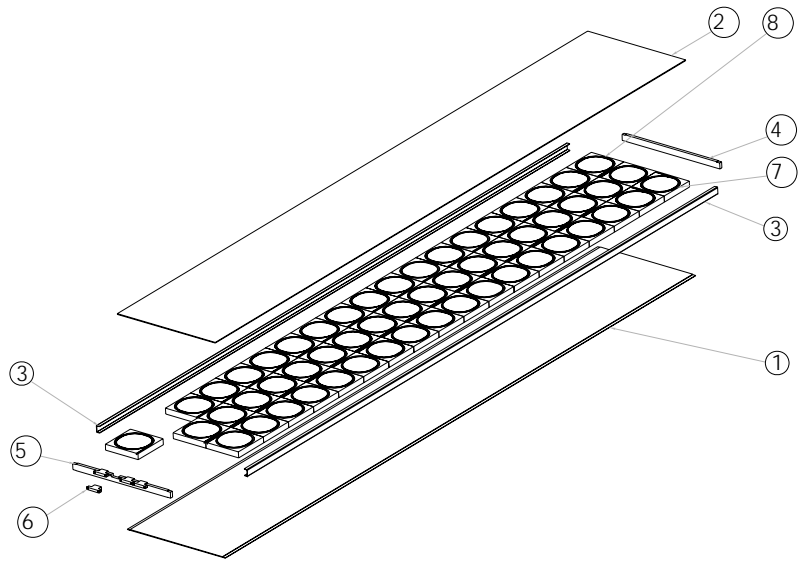

Figure 3.1: The CPR module with the fibers not shown. 1-2) Top/bottom aluminum covers, 3) aluminum U-channels, 4) PVC back piece, 5) PVC connector holder, 6) example connector for fibers, 7) scintillator tile, 8) groove for fibers.

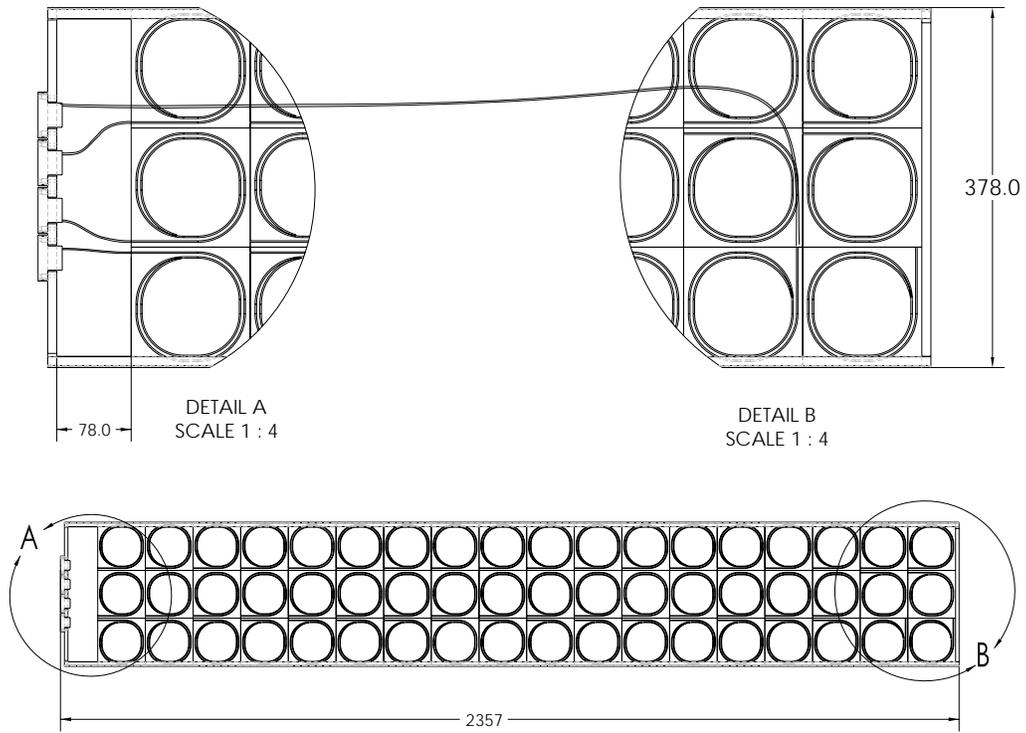

Figure 3.2: Top view of the CPR module, with only 1 fiber shown per connector for clarity.

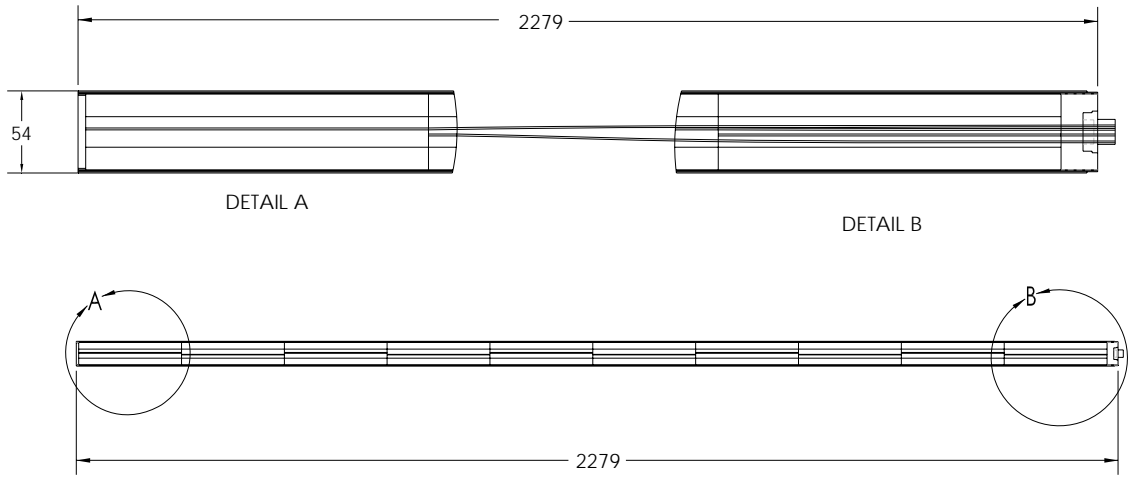

Figure 3.3: Top view of the CCR module, with only 2 fibers shown for clarity.

## 4. Preshower Scintillator Production

The scintillator tiles cutting and polishing was performed at JINR[9]. The total module width (determined by the size of three tiles, each nominally 125mm) was specified to be within 3mm of the nominal width, and the initial lot of tiles was slightly larger than this on average. This was corrected in the $2^{nd}$ lot and subsequent lots remained with the specification.

The tile edge polishing was performed in two steps. Twenty tiles were polished together using a clamp-holder fixture, and a commercial grinding device. The device was modified to provide continuous feeding of polishing suspension (the aqueous solution of the polishing powder) to the active zone. If necessary, the tile edges were polished by hand. The total polishing time for 20 tiles was 20 to 30 minutes.

In the final assembly, each tile's light yield was calibrated with a radioactive source, while a smaller sample was tested with cosmic-ray muons. As part of the tile production, a comparison was made of the light yield from the radioactive source $Sr^{90}$ and that from cosmic ray muons. The source radiation was collimated to a 1 mm diameter beam, perpendicular to the large surface of the tile (Figure 4.2). The comparison in Figure 4.3 with muon peaks using a cosmic ray trigger shows a good correlation in light yield.

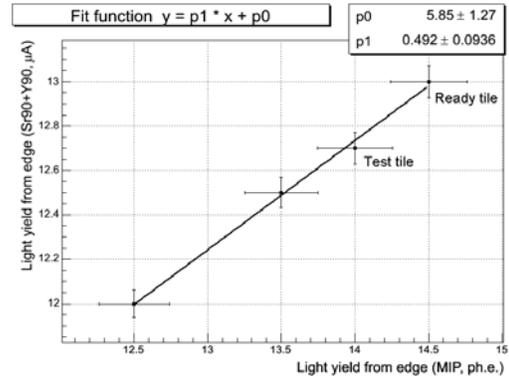

Figure 4.3: Correlation of the light yields measured by different methods.

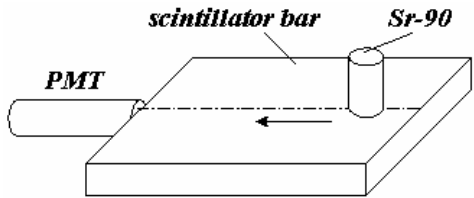

Figure 4.2: Schematic view of the source measurement setup.

## 5. Module Optical Fiber Production

As described previously, a 1.1 mm diameter multi-clad wavelength shifting fiber is inserted in each scintillator tile in a keyhole groove machined in the top surface of the scintillator. In order to optimize the amount of light transmitted to the optical connectors mounted at the end of the CPR module, most of the WLS fibers were spliced to a clear Kuraray multi-clad fiber using the thermal splicing technique first developed for the CDF endplug upgrade. The placement of a small section of Teflon heat shrink tubing over the splice allows for an increased strength at the splice position. The average transmission through the splice is of the order of 90% with an rms of a few percent.

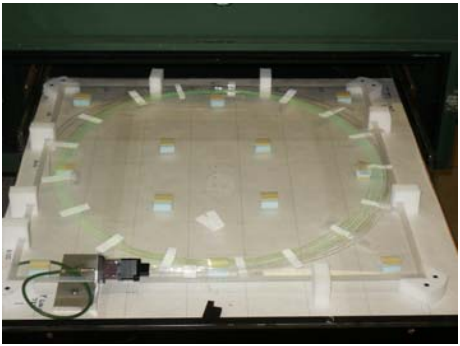

Figure 5.1: Module optical fiber tester.

The end of the WLS fiber opposite the splice has an aluminum mirror of ~3000 Angstrom thickness sputtered on using a 3 inch magnetron sputtering gun and a 99.999% chemically pure aluminum target. A protective coating of a UV epoxy manufactured by Red Spot was applied to protect the mirrored end. Both ends of the wavelength shifting fiber and one end of the clear fibers were polished using an ice-polishing technique. The wavelength-shifting clear fiber assemblies were glued (using Bicron BC-600 epoxy) into a Delrin connector to form a pigtail. After allowing for a curing time of 3 days, the connector faces were polished using a diamond-tipped flycutter. The pigtails were visually checked by scanning them with a UV light source before insertion into the CPR/CCR module.

## 6. Optical Cables

The light signals were transmitted from the CPR and CCR modules to the PMT's mounted on the backs of the calorimeter wedges using multi-clad polystyrene optical fibers similar to those used in the module pigtails. Fibers from two manufacturers, Kuraray and Polihitech, were used. The fibers from the two companies had similar optical properties. The Kuraray fibers were more flexible and were used in positions where the potential stress on the fibers was greater. The cables were routed from the high-z ends of the modules through the 2 cm crack between the central calorimeter arches and the endwall calorimeter. The fibers varied in length from 3.35 m to 4.58 m and were 1.2 mm in diameter in order to efficiently capture the light coming from the 1.1 mm diameter pigtail fibers. Each optical cable consisted of 16 fibers to match the number of pixels in the PMT's. Three of the cables from each wedge read out CPR channels while the fourth cable was split, with 6 of the channels being connected to the CPR and 10 to the CCR.

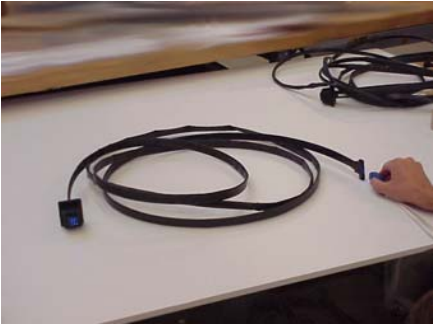

Figure 6.1: Optical cable being tested.

The optical cables were covered by a heat shrink tubing to provide both physical protection and light-tightness. The fibers were free to move with respect to each other inside the sheath so that the cable could be flexed in either direction without additional stress on the fibers. At the detector end, the optical connectors mated to similar size and shape connectors on the pigtails. The other end of the optical cable terminated in a cookie that positioned the fibers onto the pixels on the PMT's. The fibers were glued into the connectors using Bicron BC-600 optical epoxy. After a 72-hour cure the mating surfaces of the connectors were polished with a diamond cutter. Finally, the fibers were potted into black connectors at each end in order to provide additional strain relief (and light-tightness).

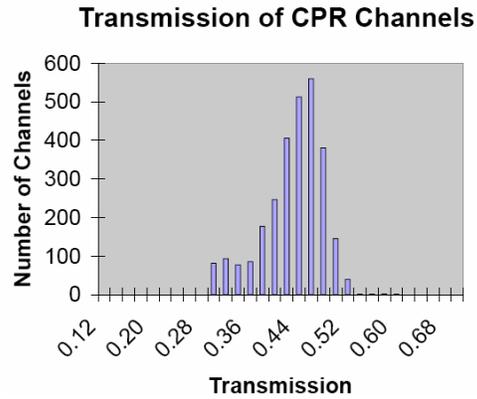

Figure 6.2: Transmission of optical cables.

The cables were tested by measuring for each fiber the transmission of light from a green LED onto a photodiode. The light was injected from a short cable through a connector identical to the one on the pigtail fibers and was compared to the light measured on the photodiode with the test cable removed. The average light transmission was measured to be 41% with a standard deviation of 4%. Cables with any channel with an average transmission of less than 70% or with RMS larger than 20% were flagged as bad and the cable was either repaired or replaced.

## 7. Photomultiplier Case

The photomultiplier tubes, described in the next section, are mounted in boxes that are located on the outside surfaces of the calorimeter wedges. The design of the boxes is shown in Figure 7.1. The optical cables are inserted into holes on one surface of the box and held securely with cap screws. A plastic mating ring fits over the optical cable end and aligns the photocathode of the PMT in a fixed orientation at a distance of .010 inches from the polished ends of the fibers. The PMTs are mounted inside of magnetic shields consisting of two concentric tubes, an inner tube made of mu-metal, .040 inches thick, and an outer tube of soft iron, .032 thick. This combination reduces the residual magnetic field at the location of the PMTs to less than 10 Gauss.

The PMTs are held in the magnetic shields by spring clips that maintain the contact with the mating rings for any orientation of the box. The side surfaces of the box contain the feed-throughs for HV, 16 anode signals and one dynode signal for each PMT. Cables and connectors inside the box allow any tube to be replaced without disconnecting external cabling. The feed throughs are light tight and electronically isolated so that the box is at the same potential as the HV ground. A lid on each box is held in place by a wing nut that allows easy access.

.

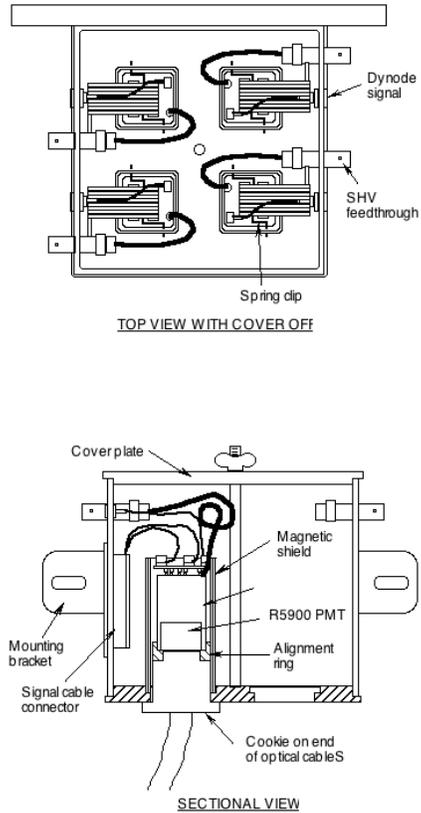

Figure 7.1: PMT box diagram.

## 8. Photomultiplier Tubes

**Overview** The light produced in scintillating tiles of the CPR2 and CCR detectors are absorbed and converted to a longer wavelength (green light) by the wavelength-shifting (WLS) fiber (Kuraray Y11) imbedded in the groove in the tile. The green light is then transmitted through a clear fiber to `photomultiplier` tubes (PMTs) located outside of the calorimeter arch.

The total number of channels is 64x48=3072. Because of cost and space limitations, it would have been impractical to read out individual channels using conventional, single-channel PMTs. Therefore we have chosen to use multi-anode photomultiplier tubes (MA-PMTs), which employ a metal-channel dynode structure and provide multiple anode readout channels within a single physical tube. Hamamatsu Photonics produces such PMTs, with several different varieties in the pixel structure with a common design in the overall size of the tube envelope. We have chosen a 16-channel version, which is shown schematically in Figure 8.1. The size of the pixels is 4 mm x 4 mm. Some modifications (to be described later) have been applied to the original version, and the model number of H8711-10MOD2 is given to our version. It has a normal, bialkaline photocathode, sensitive mainly to blue light and 12-stage metal-channel dynodes and a 16-channel anode. The model also has an output from the final dynode stage, DY12.

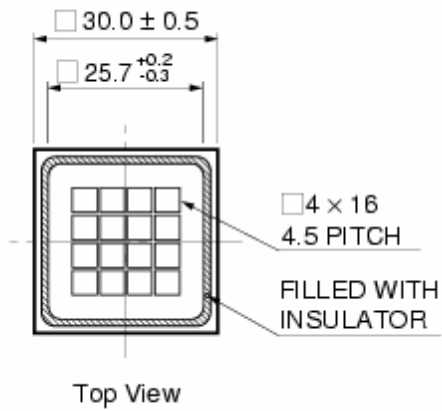

Figure 8.1: PMT top view.

The high voltage bleeder circuit for H8711-10 has non-equal voltage division among multiplication stages ("tapered" base) for a better linearity for short pulses. A typical gain at 800 V is $10^6$. The cathode and the dynodes are electrically common to all 16 channels and only the anode has the pixel structure. Dividing the total channel count of 3072 by the number of channels per tube, 16, gives 192, which is the number of tubes we use for the CPR2 detector.

We have tested the PMTs both during the R&D and mass production periods. In the remainder of this Section we describe them in more detail. To be more specific, we have examined if (a) the tube has a sensitivity to the expected MIP signal (5 to 10 photoelectrons) with good pulse height resolution, (b) the response variation among 16 pixels within a tube is small enough, (c) the cross-talk among anode pixels, which is inevitable by tube construction, is small enough, and (d) the linearity is still good for larger pulses expected from high energy electromagnetic showers. Below we describe measurements we have performed related to these characteristics.

**Mechanical and Electrical Structure**

The size of the PMT assembly is 30 mm x 30 mm x 45 mm, as shown in Figure 8.1. The sensitive cathode area is 18.1 mm x 18.1 mm. The schematic diagram of the electrical circuit, including the high voltage bleeder, is shown in Figure 8.2. The cathode is at the negative voltage with respect to the ground. The voltage division employs a ratio of 1.5: 1.5: 1.5: 1: 1: ...: 1: 2: 3.6. It is called a "tapered" base, and is intended to provide a better pulse linearity. A non-linearity is caused mainly by space-charge effects at the final dynode stages if a large number of multiplied electrons reside within a small physical region. This is eased by providing higher voltage gradients at those stages and forcing the electrons to move quickly to the next dynode stage and to the anode.

The following modifications are made to the original design of the PMT. The original version of H8711-10 had 100 kΩ resisters between the individual anode channels and the ground. They create small DC offsets to the anode signal lines and had been known to cause problems to the existing CDF electronics. Therefore they have been removed. Also, a register, R19, has been added to the circuit in order to prevent a possible ground loop. The high voltage is provided through a red, RG-174/U coaxial cable. In the original version the anode output was provided by individual thin coaxial cables.

In the CPR2 detector design, we use short, mass-terminated flat cables within the PMT box to interface the anode signals at the wall of the box to the electronics outside. For this purpose a modification is made so as to provide the anode signals with pins. The pins are SAMTEC TSW-108-07-T-D, which mate with standard 0.1-inch pitch flat cable connectors such as SAMTEC IDSD-08. The definition of the channel numbers is given as follows. If you look at the tube into the cathode (top view), the top-right corner is Channel 1, the top-left is Channel 4, and the bottom-left corner is Channel 16 (Figure 8.3). They are engraved in the piece of metal near the tube surface.

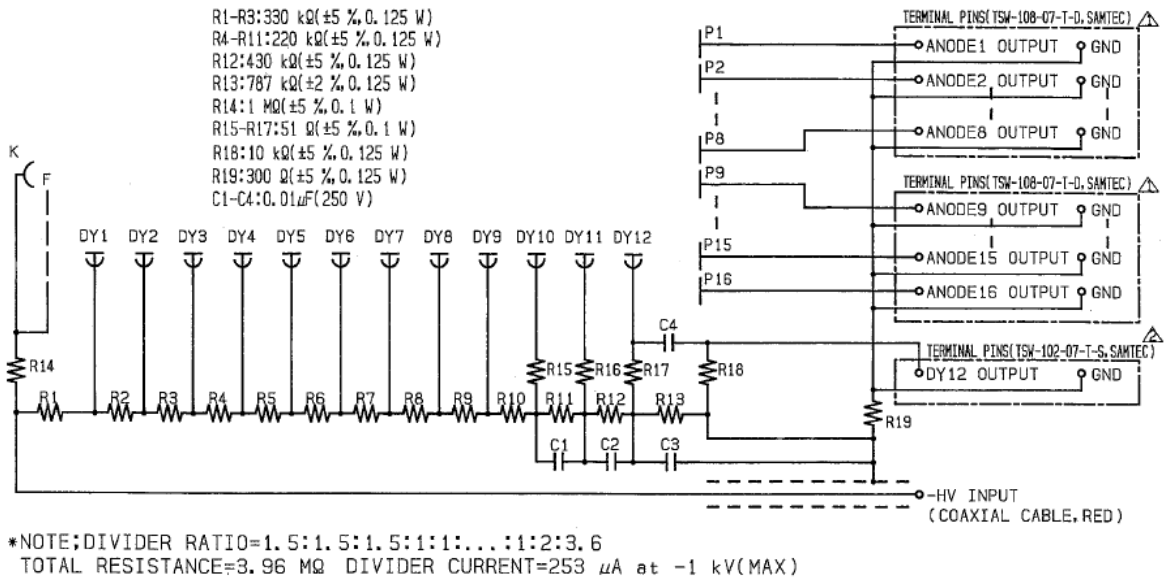

Figure 8.2: PMT base schematic.

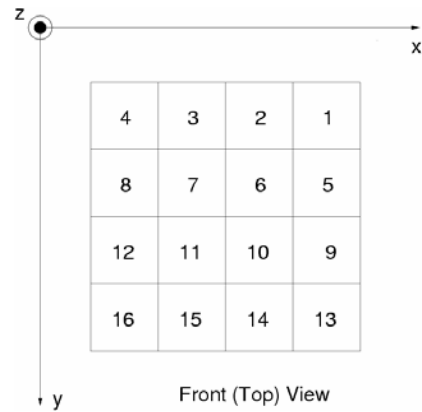

Figure 8.3: Coordinate system.

**Test measurement setup**

In this section we describe the experimental setup that is used to measure and examine the characteristics of the PMTs. We use the Hamamatsu diode laser PLP-02/041 as our light source. It consists of a precision pulser and a 405-nm diode laser head. It provides short pulsed light with stability and low time jitter.

To test the PMTs with conditions similar to those in the real detector, the blue light is guided through a quartz fiber-optic cable (core diameter 50 mm) to a Kuraray Y-11 wavelength-shifting fiber (0.83 mm in diameter). The blue light is converted in the fiber to a longer wavelength (green), and eventually reaches the face of the MA-PMT. There is an air gap of 1 - 2 mm between them. The other end of the WLS fiber goes to a conventional PMT (Hamamatsu H1161 with green-extended photocathode) for monitoring purposes.

The MA-PMT is placed on a two-dimensional scanner so that effectively the fiber scans over the surface of the MA-PMT for uniformity measurements. The anode signals of the PMTs are measured with a CAMAC ADC (LeCroy 2249W, nominal conversion factor of 0.25 pC per count) and read out through a PC. The ADC gate width is set to 150 ns. The scanner movement is generated by pulse stepping motors and is controlled through the PC as well. The definition of our coordinate system is given in Figure 8.3.

The CPR2 detector is a device intended to measure energy deposits by particles. Therefore, it is important to have responses that are as uniform as possible. The PMT gains among individual tubes can be equalized by applying different high voltages to different tubes. However, the high voltage is common to all 16 channels within a tube, and we cannot adjust gains among those 16 channels. They can be measured beforehand and can be corrected by applying those measurements as calibration constants.

Hamamatsu provides their own measurement of anode uniformity among 16 channels in each tube, as part of the data sheet, along with other data such as cathode and anode sensitivities. However, it turns out that the uniformity depends on the way the PMT is used, and we performed our own measurements of the uniformity. The uniformity provided by Hamamatsu is a relative one, normalized to the channel of the highest response (which is defined to be 100). For the purpose of comparison, we have re-normalized them to the average of the 16 channels, which is defined to be unity.

If we examine the distributions further by dividing the channels into four regions within a tube, one can see that the four "corner" channels (Channels 1, 4, 12, 16) tend to have the highest response, and the four channels in the "middle" (Channels 6, 7, 10, 11) have the lowest response. We shall see that this is not a coincidence and does reflect tube construction.

**Uniformity at the center of the pixels**

The design of the CPR2 detector calls for one and only one WLS fiber per detector channel. With this design the fibers will naturally be guided to the centers of the 16 pixels of the PMT. Therefore we want to know the tube response at those 16 points on the PMT.

For this we use the setup described earlier, where the PMT moves with respect to the light source and we can scan and measure responses over the surface of the PMT. We measure the response at the 16 points, corresponding to the centers of the pixels, and then normalize them to the average of the 16 responses. The width of the distribution provides a measure of the response uniformity, which is 12.5%.

The number, 12.5%, is different from the Hamamatsu number, 17.9%. If we assume the two sets of measurements are both correct, or at least not grossly wrong, there has to be a reason for a systematic difference. An indication that there indeed exists such a reason can be seen in Figure 8.4, where a comparison of two sets of measurements is given separately for different regions in the PMT. One can see that overall correlations exist between the two sets. However, for the "corner" channels, Tsukuba measurements tend to be lower than Hamamatsu measurements, while for the "middle" channels it is vice versa. For the channels in the "top/bottom" (Channels 2, 3, 14, 15) and "left/right" (Channels 5, 8, 9, 12) regions, the discrepancies are smaller. In the next subsection we will describe a reason why it happens.

**Anode uniformity fine scan**

In order to understand the response uniformity over the entire surface of the PMT, we have taken data by scanning the PMT at 0.5 mm intervals. An example of such a measurement is shown in Figure 8.5. The response of a single channel (Channel 1) is shown as a function of the position (x, y) of the illumination point. The colored bands (25% intervals) are relative to the highest response point. The coordinate system was defined in Figure 8.3. The origin does not coincide with an edge of the tube.

From such two-dimensional maps, one can extract one-dimensional slices, the response along one of the axes (e.g. x) while keeping the position on the other axis near the center of the pixels. This is shown in Figure 8.6. Solid vertical lines corre-spond to the pixel boundaries, and dashed lines the centers. Squares show the response of the reference tube. The data points in these one-dimensional slices are subsets of the two-dimensional fine-scan data.

One may notice that the regions of sensitivity extend beyond the nominal pixel boundaries toward outside, notably towards the outside of the tube. This is not a characteristic of this particular tube, rather is common to all tubes. It seems likely that the behavior comes from tube construction.

Now, we take the responses such as in Figure 8.6 and add them all up. This "integrated response" would be the response we would observe on this channel if we used a light source located sufficiently away from the tube and shone the entire surface of the cathode uniformly ("full illumination"). If we define this integrated channel responses of individual channels from fine scan data as the channel response and compare them with the Hamamatsu anode uniformity measurements, we obtain what is shown in Figure 8.7. Data from nine tubes (R&D version) are shown. It is to be compared with Figure 8.4.

Now the agreement is much better, and in all regions the points tend to lie much closer to the line with y = x. And it has turned out that it is indeed anode uniformity. So we were comparing apples and oranges in Figure 8.4.

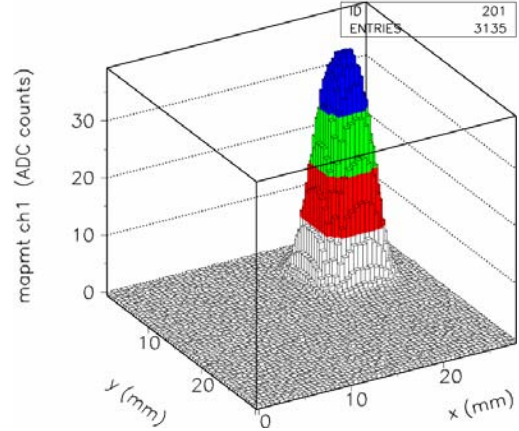

Figure 8.5: Anode uniformity fine scan described in text.

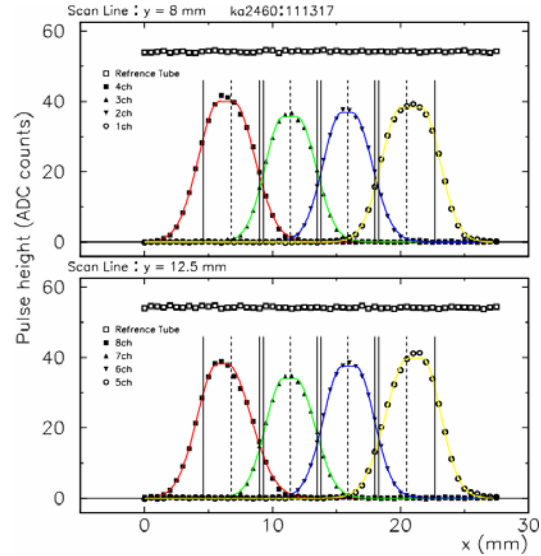

Figure 8.6: Response of four channels in a given row of a PMT as a function of the illumination position x while keeping the position y near the center of the pixels.

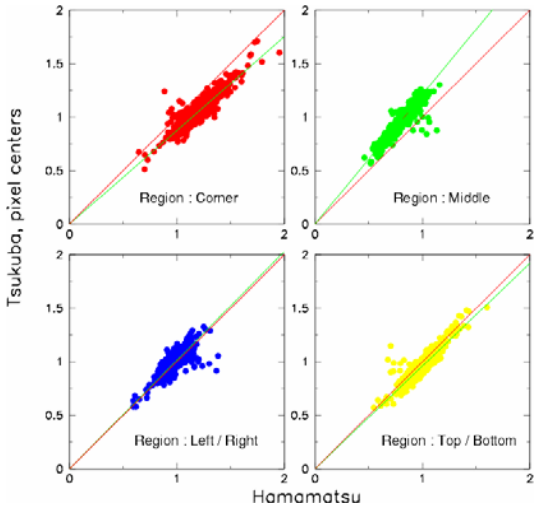

Figure 8.4: Comparison between Tsukuba measurements of responses at the center of 16 channels with that from Hamamatsu anode uniformity data, for 200 PMTs. The four subplots correspond to the four regions of the PMT.

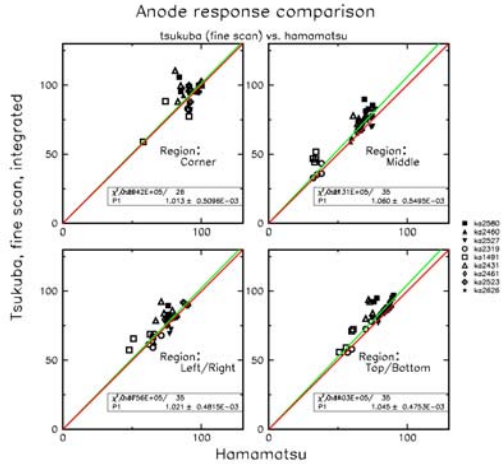

Figure 8.7: Comparison of the integrated anode responses from the Tsukuba fine-scan to the Hamamatsu anode uniformity data.

**Cross talk**

The multi-anode PMT enables us to reduce the number of tubes and thus the cost, while keeping a desired fine segmentation of the detector. However, some cross talk among the channels within a tube is inevitable because of the tube construction.

The Hamamatsu catalog lists a cross talk at the per cent level. Our data with the light source at the center of the 16 channels also provide cross talk measurements, because we read out all 16 channels of the tube. The response observed at the direct neighbors (those sharing the sides) is at the 2 to 3% level of the pixel with the light source, while the cross talk to diagonal neighbors is around 0.5%. The amount of cross talk depends on how we guide the fibers to the photocathode. The light exiting the fiber has a finite cone angle that is determined by the index of refraction of fiber material. Therefore, the distance from the edge of the fiber to the tube envelope matters.

In our measurement setup, we have to have a finite space between the fiber and the envelope, because we have to move the tube relative to the fiber without touching each other. That distance is about 1 to 2 mm. In the real CPR2 detector the distance will be almost zero. Therefore the cross talk numbers given here are an upper limit.

**Pulse linearity**

It has been known that multi-anode PMTs with small physical sizes show non-linear responses at large anode signals. This is caused by space-charge effect at the final few stages of the dynodes and the anode. The effect is known to be alleviated by using "tapered" bases, as opposed to "linear" bases. The tapered bases provide higher voltages at these final stages and force the electron cloud to move quickly to the next stage of the dynodes and to the anode. The PMT with this tapered base circuit is given the name H8711-10, with "10" meaning a factor of ten better dynamic range. Hamamatsu specifies the pulse linearity as the point where the anode output deviates from a linear behavior by 2%. For H8711 tubes, the 2% point corresponds to anode currents of 0.5 mA and 5.0 mA for the linear and tapered bases, respectively.

We want to think in terms of output charge, since we are dealing with pulsed signals instead of constant direct currents. The shape of the signal from the CPR2 detector is determined by the decay time of the Y-11 wavelength shifter, which is of order 10 ns. It can be dispersed during transmission through long clear fibers.

If we operate the tube at a gain of $10^6$, a 10 photoelectron MIP signal will produce a charge of 1.6 pC. If this amount of charge is collected by the tube with a time duration of 10 ns, the average current will be $I_{pulse}$ = 1.6 pC /10 ns = 0.16 mA. So the 2% saturation point is only at 3 MIPs with a linear base and at 30 MIPs with a tapered base.

We have measured the pulse linearity with our own setup. We do use light from a Y-11 WLS fiber, so it should be close to what we have in the real detector. We have measured output charge by applying different high voltage values to the tube. It is known that the gain changes with a certain power of the HV value. If we plot charge vs. HV on log-log scales, it should lie on a straight line with the absence of non-linearity, and a deviation toward a lower output would indicate saturations.

Figure 8.8 shows results of such a measurement for three tubes, KA2580, KA2626 and KA2418. The first two have a linear base, and KA2418 has a tapered base. Two sets of measurements exist for each tube, performed with different amounts of input light, roughly 10 and 1000 photoelectrons. One can easily notice that the tapered base has a much better linearity than the linear base, although the gains are lower. The linear-base tubes seem already saturated at 800 V, at which point the tube gain (from Hamamatsu data sheet) is $5.1 \times 10^6$ and $7.0 \times 10^6$.

Another thing one may notice is that the saturation seems to occur at different output charges depending on the amount of input light. One may naively expect that the saturation occurs at a fixed output charge. However, it is not strictly true and can be explained as follows. The peak anode current depends on the time structure of the pulse, even if the total charge of the pulse is fixed. The time structure reflect the times when individual photons are produced in Y-11 wavelength shifting fibers, and thus on the number of photoelectrons produced at the cathode.

For an extreme case of single photoelectrons, the time structure is a delta function. On the other hand, if there are a large number of photoelectrons, they are a collection of photons produced at different times. The pulses amplified in the PMT reflect these time structures of parent photons, and the single photoelectron pulses will have a very narrow width and thus a higher peak current. In the detector large pulses are the result of higher energy deposits and thus a larger number of photons. From the previous figure, we find that pulses as large as 50 pC can be measured reasonably accurately with the PMTs having the tapered base.

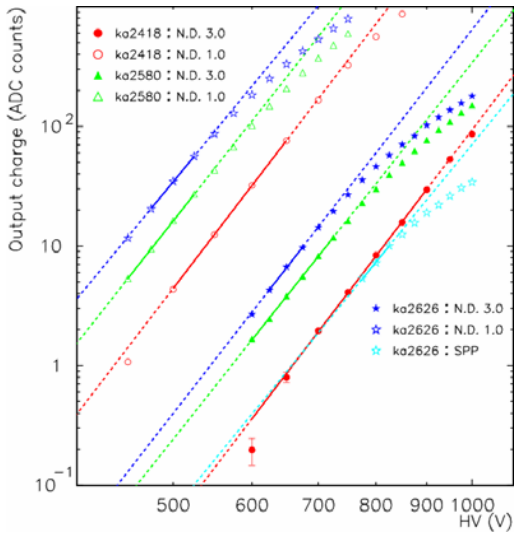

Figure 8.8: Measurements of PMT output versus applied high voltage.

**Photoelectron peaks and the absolute gain**

It is sometimes desirable to know the absolute gains of PMTs, not just relative ones. The absolute gains can be determined if one can observe single photoelectrons, which requires high gains so that the peak can be separated from pedestals.

H8711 multi-anode PMT can be applied a high voltage of as high as 1000 V. Figure 8.9 shows an example of pulse height distributions taken with a small amount of injected light (roughly 0.1 photoelectrons). The single photoelectron peaks are clearly visible.

Hamamatsu provides cathode and anode sensitivities (called $S_K$ and $S_P$, respectively) in data sheets. We can calculate the gain from the two quantities by Gain = $S_P/S_K$. They are measured at a high voltage of 800 V. To make comparisons, the gains measured from single photoelectron peaks at 1000 V are converted to the corresponding numbers at 800 V, using a measured dependence of the anode response on the applied high voltage value. As described earlier, we can use a relation $Q/Q_0 = (V/V_0)$ as a good approximation, where $Q$ ($Q_0$) is the output charge at voltage $V$ ($V_0$), with being a constant around 10.5.

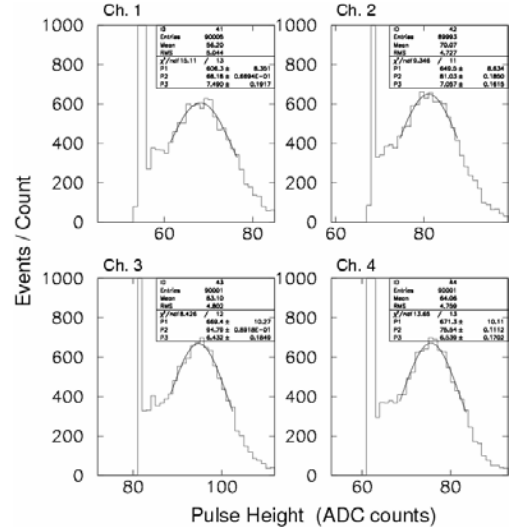

Figure 8.9: Pulse height distributions with very low light level (approximately 0.1 photoelectrons)

The distributions and gains in Figure 8.9 are for individual channels of a tube. We correct them for the anode response of the channels, and then calculate one gain number for the tube. They are then compared with the Hamamatsu numbers and are shown in Figure 8.10. The Tsukuba measurements give higher gains than Hamamatsu. The ratio of gains (Tsukuba to Hamamatsu) has an average value of 1.60 with rms spread of 0.19 for 11 tubes measured.

The difference is partially traced to the way they are measured. The single photoelectron method measures the amplification of single photoelectrons that arrived at the first stage of the dynodes. On the other hand, the "gain" we calculate from the Hamamatsu measurements are from the cathode and anode currents. The cathode current is a measure of the number of photoelectrons that are produced at and left the cathode, but they do not necessarily arrive at the first dynode. The "gain" defined as the ratio of those anodes to cathode currents will be smaller than that by the single photoelectron measurements. The probability of arrival at the first dynode is called collection efficiency. It depends on the voltage applied between the cathode and the first dynode, and plateaus at about 80% for a sufficiently high voltage.

The observed gain ratio of 1.60 implies an efficiency of 63%, which is somewhat lower than expected. We operate the PMTs in the CPR2 detector with a gain of $10^6$ by adjusting high voltage values to individual tubes. We assume a factor of 1.6 when we calculate the necessary high voltage.

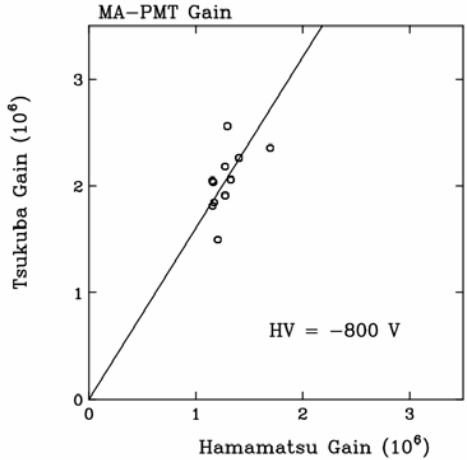

Figure 8.10: Comparison of PMT absolute gains measured at Tsukuba with those from Hamamatsu data sheet values.

**Effect of the magnetic field**

It is a well-known fact that PMTs in general are susceptible to operations in magnetic fields. The electrons can suffer changes in trajectories and thus can fail to reach next stages of multiplication, resulting in loss of gain.

The CPR2 detector houses the PMTs in mu-metal shields and at the location of the PMT boxes the residual magnetic field from the CDF solenoid is expected to be of order 1 Gauss.

Hamamatsu measurements had shown the effects depend on the direction of the magnetic field. In the coordinate system defined in Figure 8.3, a field in the positive y direction had given the largest effect, resulting in a gain loss of about 3%, 12%, 23%, 35% and 50% for fields of 10, 20, 30, 40 and 50 Gauss, respectively. This was measured with a 64-pixel version of the MA-PMT. Therefore, we do not expect significant effects from the field in practice.

Nevertheless, we have examined the effects of the magnetic fields on the MA-PMTs using small pieces of permanent magnets. They produced magnetic fields of 30 – 40 Gauss in x and y directions. We have observed that the anode output decreased by 5% to 10% among 16 channels when a field existed in the positive y direction. For the other three directions, the effects are almost unnoticeable, at most 2% level. For the z directions, the magnets produced fields of only up to about 5 Gauss, and no significant effects are observed, as expected.

These results are fully compatible with Hamamatsu measurements, and we should not expect any deterioration in the performance for the use in the CPR2 detector. Also, the cross talks showed no difference from a measurement when a field was absent.

## 9. High Voltage

The Hamamatsu R5900 PMT's operate at a gain of $\sim 10^5$, at voltages between 500 and 600 V, with a tapered bleeder which offers a good linearity as a function of the anode peak current. There are 30 high voltage (HV) channels for the CCR and 162 HV channels in the CPR detector, for a total of 192 HV channels, supplied by a CAEN SY527 crate lodging eight HV boards CAEN A932 AN. Each board fans out the HV to as many as 24 channels: each channel can be independently regulated and monitored within a range of 900 Volts, starting from the input voltage down.

Each supply channel has a programmable voltage setting, programmable fast current trip, trip on power and an overvoltage protection trip. All the HV channels are controlled by a PC that ramps all the supplies on equally, in proportion to their final value. The HV control software provides for software current trips and ramping (up and down) of either individual channels or groups corresponding to a calorimeter arch.

Each twelve outputs of the A932 AN card are carried, through an AMP connector, by a Silisol multiconductor HV cable to the distribution box mounted on each calorimeter arch. Short RG-58 HV cables then go from the distribution box to each phototube on the back of the calorimeter wedges.

## 10. Electronics, Assembly, Installation

The electronics for both CPR and CCR detectors re-used the shower maximum electronics system described in reference [10]. Formerly unused channels, 22 per module, were activated for the upgrade. The pre-amplifier boards needed for the old gas-based detectors were replaced with simpler feed-thru boards appropriate for PMTs.

Detector module assembly proceeded as follows. Module fibers were prepared as discussed in section 5, and then inserted into the tiles already positioned on the module bottom. The sides and top covers were then glued into place. Each assembled module was then scanned with an automated $^{137}$Cs source system. Figure 10.1 shows the distribution of tile gains, the width of the distribution was due to scintillator variations. These gains were stored in a database and used as the first iteration of the detector calibration.

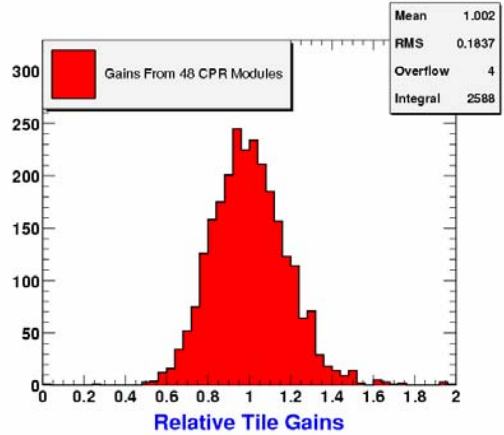

Figure 10.1: Distribution of relative tile gains from a radioactive source scan.

The detector was installed in the CDF collision hall, with a very constrained access. All the personnel and equipment passed through an 18" x 13" hole, then on the inside of the detector a 3-story scaffold was assembled to allow installation. The installation was successfully completed in a twelve week period with 99.9% of the detector channels working (3 bad channels out of 2592). A schematic of the installation scaffold is shown in Figure 10.2.

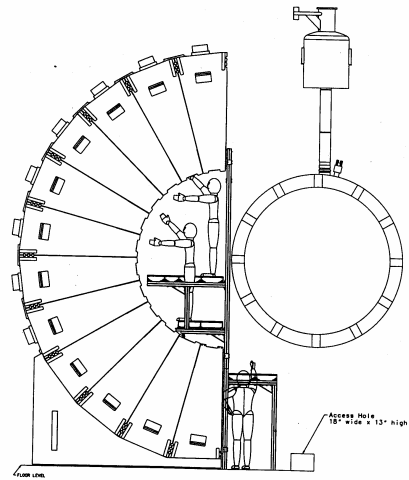

Figure 10.2: Installation configuration. Box on lower right was the access hole.

## 11. Results from Collider Data

The CPR and CCR detectors have collected a large amount of proton-antiproton data since installation. The individual CPR tile gains have been calibrated using reconstructed tracks that are extrapolated to the tile. More than 99.8% of the CPR channels are functional and calibrated with this technique. Minimum-ionizing peaks for each tile are recorded and used to correct the data in physics analyses. Figure 11.1 shows the data distribution for muon candidates from W boson decays, after correction for tile gains, in terms of equivalent number of minimum-ionizing particles. For comparison the distribution of random tracks with $p_t$ greater than 8 GeV, which sometimes produce particle showers in the material in front of the CPR, is also shown.

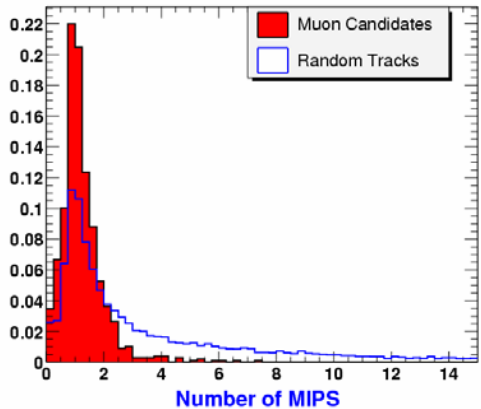

Figure 11.1: Number of equivalent minimum-ionizing particles is shown for muon candidates from W boson decays, and also from random tracks.

In addition to MIPs, many electrons from W boson decay have been recorded in the CPR detector. Figure 11.2 shows the distribution of pulse height from these electron candidates, compared to the same random track distribution from the previous figure 11.1. As expected for electron showers, the pulse height is larger than single MIPs and random tracks.

The CCR detector sits behind 9.6 X0 of material in the solenoid magnet material plus a tungsten bar. It therefore is designed to observe electromagnetic showers, not minimum-ionizing particles. Calibration with proton-antiproton data is best performed with Z boson decays into electron-positron pairs, with one leg striking the crack detector. The electron-pair mass is reconstructed with the other electron plus the reconstructed track 4-vector that strikes the CCR detector, and is required to be consistent with a Z boson. The CCR pulse height is calibrated with the track momentum. Several wedges were observed to have lower pulse height than the rest of the detector, and more pulse height sharing between tiles. This was attributed to poor optical connections between the detector and optical cable when installed. After calibration, however, these wedges perform well as electromagnetic shower detectors.

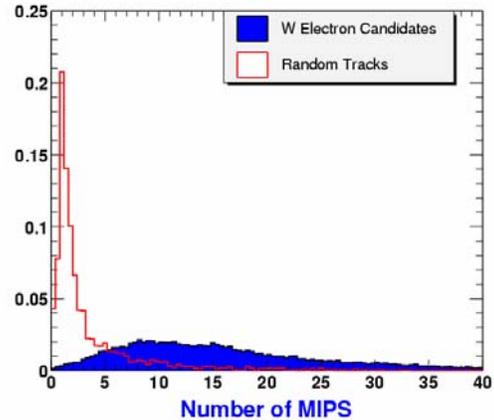

Figure 11.2: CPR pulse height distribution for W electron candidates compared to random tracks.

After calibration constants are applied, the CCR energy is compared to the track $p_t$, in Figure 11.3. Good correlation is observed. The non-linearity observed at higher energies is expected due to the change in shower profile with energy. An additional source of non-linearity affecting the curve is the saturation of some channels of the electronics. This will not affect the physics analyses that use the CCR to identify electromagnetic showers hitting the crack detector.

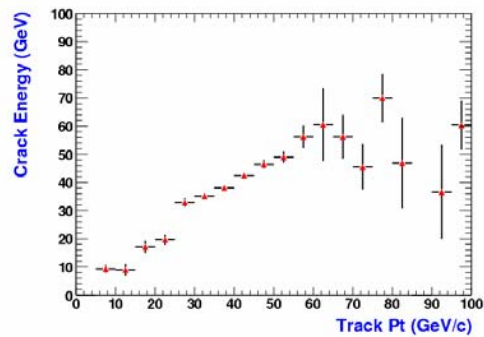

Figure 11.3: Crack detector energy response versus track $p_t$ from Z boson electrons.

## 12. Acknowledgements

We acknowledge the contributions of the Italian and Korean groups, who provided more than 25 technicians and scientists for the module assembly at Argonne National Laboratory. The work of Mike Nila and his staff at Michigan State University was crucial in delivering the prepared fibers and optical cables necessary for installation. The technical staff at Argonne, led by the engineering of Jim Grudzinski and technician Zeljko Matijas, made vital contributions in module assembly and testing. Finally, the challenging installation was led by Fermilab's Stefano Moccia who designed the 3-story scaffold. Other major contributions from FNAL staff came from Dervin Allen, Lew Morris, Jim Humbert, Craig Olsen, Dean Beckner, and Luda Rychenkov.